\documentclass[10pt]{iopart}
\usepackage{epsfig,cite}
\newcommand{\ewxy}[2]{\setlength{\epsfxsize}{#2}\epsfbox[30 30 640 640]{#1}}
\newcommand{\beq}{\begin{equation}}
\newcommand{\eeq}{\end{equation}}

\newcommand{\beqar}{\begin{eqnarray}}
\newcommand{\eeqar}{\end{eqnarray}}

\begin{document}

\title[Resonance-rich matter in HICs at RHIC]{Transition to 
resonance-rich matter in heavy ion collisions at RHIC energies}
\author{
L~V~Bravina\dag\ddag, E~E~Zabrodin\dag\ddag,
M~Bleicher\S, S~A~Bass$\Vert$, M~Brandstetter\P, 
A~Faessler\dag, C~Fuchs\dag, W~Greiner\P,
M~I~Gorenstein$^+$,
S~Soff$*$,
H~St\"ocker\P
}
\address{\dag\
Institut f\"ur Theoretische Physik, Universit\"at
    T\"ubingen, T\"ubingen, Germany}
\address{\ddag\
Institute for Nuclear Physics, Moscow State University, Moscow, 
    Russia}
\address{\S\
Nuclear Science Division, Lawrence Berkeley Laboratory, USA}
\address{$\Vert$\
National Superconducting Cyclotron Lab, Michigan State 
    University, USA}
\address{\P\
Institut f\"ur Theoretische Physik, Universit\"at Frankfurt, 
    Frankfurt, Germany}
\address{$^+$\
Bogolyubov Institute for Theoretical Physics, Kiev, Ukraine}
\address{$*$\
Gesellschaft f\"ur Schwerionenforschung, Darmstadt, Germany}

\begin{abstract}
The equilibration of hot and dense nuclear matter produced in the
central region in central Au+Au collisions at $\sqrt{s}=200${\it A\/} 
GeV is studied within the microscopic transport model UrQMD.
The pressure here becomes isotropic at $t \approx 5$ fm/$c$. Within 
the next 15 fm/$c$ the expansion of the matter proceeds almost 
isentropically with the entropy per baryon ratio $S/A \approx 150$. 
During this period the equation of state in the 
$(P,\varepsilon)$-plane has a very simple form, 
$P=0.15\, \varepsilon$.
Comparison with the statistical model (SM) of an ideal hadron gas
reveals that the time of $\approx 20$ fm/$c$ may be too short
to attain the fully equilibrated state. Particularly, the fractions 
of resonances are overpopulated in contrast to the SM values. 
The creation of such a long-lived resonance-rich state slows down 
the relaxation to chemical equilibrium and can be detected 
experimentally. 
\end{abstract}

\section{Introduction}
The assumption that strongly interacting 
matter, produced in nucleus-nucleus collisions at high energy, can 
reach the state of local equilibrium (LE) \cite{Fer50,Land53} is one 
of the most important topics in the relativistic heavy ion programme
\cite{Tor99}.
The degree of equilibration can be checked by fitting the measured
particle yields and transverse momentum spectra to that of the thermal 
model in order to extract the conditions of the fireball at the 
chemical and thermal freeze-out \cite{BrSt96,Bec,Raf,ClRe99,YeGo99}. 
Here the equilibrium particle abundances,
which correspond to a certain temperature $T$, baryon chemical
potential $\mu_{\rm B}$, and strangeness chemical potential 
$\mu_{\rm S}$, can be determined. However, the analysis is 
complicated, e.g., by the presence of collective flow and the
non-homogeneity of the baryon charge distribution in the reaction 
volume. The volume of the fireball is also taken as a free parameter
in the thermal model. Various non-equilibrium microscopic transport 
models have been applied to verify the appearance of at least local 
equilibrium in the course of heavy ion collisions at relativistic 
and ultra-relativistic energies 
\cite{GeKa93,Cass,Bass98,Blei98,SHSX99,CBJ99}.
To reduce the number of unknown parameters and simplify the analysis 
it has been suggested \cite{lv98plb,lv99prc,rhic} to examine the 
equilibrium conditions in the central cell of relativistic heavy ion 
collisions. For this purpose the microscopic transport model UrQMD
\cite{urqmd} is employed. Previous studies at energies from 
10.7{\it A\/}~GeV (AGS) to 160{\it A\/}~GeV (SPS) 
\cite{lv98plb,lv99prc} have shown that the cubic cell with volume 
$V = 125$ fm$^3$ is well suited for the analysis. 
The aim of the present paper is to study the relaxation of hot nuclear
matter, simulated within the microscopic model, in the central cell in
Au+Au interactions at $\sqrt{s}=200${\it A\/}~GeV.

\section{Relaxation to thermal and chemical equilibrium}
\begin{figure}[htb]
\vspace{-2.8cm}
\hspace{3.cm}
\ewxy{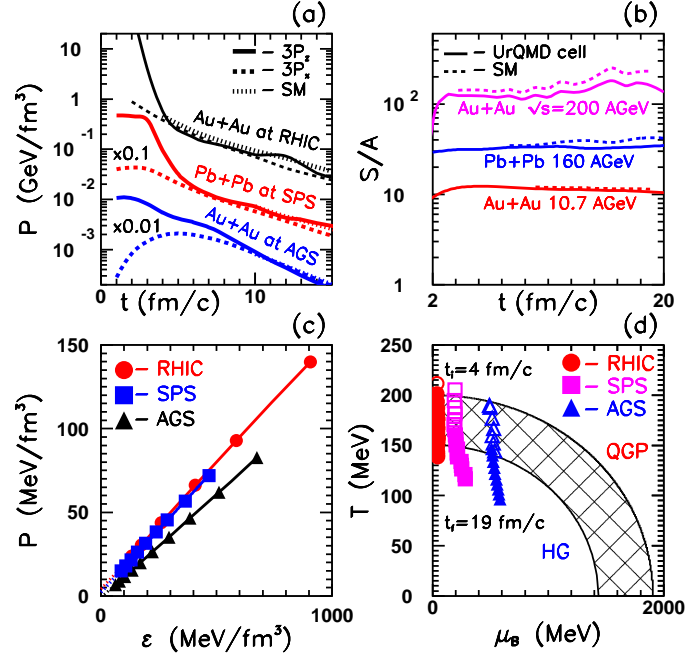}{115mm} % Combined plot of 4 figures
\vspace{-1.0cm}
\caption{
(a) The longitudinal (solid lines) and the transverse
(dashed lines) diagonal components of the pressure tensor $P$ in
the central cell of heavy ion collisions at AGS, SPS, and RHIC
energies compared to the SM results (dotted lines).
(b) Time evolution of the entropy per baryon ratio, 
$s/\rho_{\rm_B} = S/A$, in the central cell 
calculated with the UrQMD (solid lines) and the SM (dashed lines).
(c) The evolution of pressure $P$ and baryon density 
$\varepsilon$ in the central cell.
(d) The same as (c) but for the $(T, \mu_{\rm B})$-plane.
Solid symbols correspond to the stage of kinetic equilibrium,
open symbols indicate the pre-equilibrium stage.
The hatched area shows the expected region of the quark-hadron
phase transition.
}
\label{fig1}
\end{figure}

First, the kinetic equilibrium has to be verified.
The collective flow in the cell should be isotropic and small, so it
cannot significantly distort the momentum distributions of particles.
Microscopic simulations show that the longitudinal flow rapidly 
drops and converges to the developing transverse flow 
\cite{lv99prc,rhic}.
Isotropy of the velocity 
distributions results in the pressure isotropy. Pressure in 
longitudinal direction in the cell, calculated according to the virial 
theorem, is compared in figure~\ref{fig1}(a) 
with the transverse pressure. The time of convergence of longitudinal 
pressure to the transverse one decreases from 10 fm/$c$ to 5 fm/$c$ 
with rising incident energy from AGS to RHIC, respectively.
Thus, the kinetic equilibrium in the central cell in Au+Au collisions 
at RHIC energy is reached at $t \approx 5$ fm/$c$.

To verify that the matter in the cell is in thermal and chemical
equilibrium one has to compare the snapshot of hadron yields and 
energy spectra in the cell with the equilibrium spectra of hadrons
obtained in the statistical model (SM) of an ideal hadron gas.
The description of the SM employed in our calculations can be found
elsewhere \cite{Belk98}.
The values of energy density $\varepsilon$, baryon density 
$\rho_{\rm B}$, and strangeness density $\rho_{\rm S}$, extracted   
from the microscopic calculations in the cell, are used as an input
to the system of nonlinear equations of the SM \cite{Belk98} to
determine the temperature $T$, baryon chemical potential 
$\mu_{\rm B}$, and strangeness chemical potential $\mu_{\rm S}$.
This enables one to calculate the particle yields $N_i^{\rm SM}$,
total energy $E_i^{\rm SM}$, pressure $P^{\rm SM}$, and entropy
density $s$. The thermal and chemical equilibrium is assumed to 
occur in the cell when the microscopic spectra of hadrons 
become close to the spectra predicted by the SM. 

At the isotropic stage the total microscopic pressure is close to
the grand canonical pressure, $P^{\rm SM}$, as shown in 
figure~\ref{fig1}(a). Therefore, the time $t=5$ fm/$c$ is chosen as a 
starting point for the comparison with the SM. 
The final time of the 
calculations, defined from the conventional freeze-out conditions
$\varepsilon \approx 0.1$ GeV/fm$^3$ or 
$\rho_{\rm tot} \approx 0.5 \rho_0$ \cite{lv99prc}, corresponds to 
$t \approx 21$ fm/$c$. 
The hadron-string matter in the central cell expands
nearly isentropically, see figure~\ref{fig1}(b), with 
$s/\rho_{\rm B} \equiv S/A \approx 12$
(AGS), 32 (SPS), and 150 (RHIC). The results of the
simulations at AGS and SPS energies are intriguingly close to the 
entropy per baryon values extracted from the thermal model fits to 
experimental data, namely, $(S/A)^{\rm AGS} \approx 14$ and 
$(S/A)^{\rm SPS} \approx 36$ \cite{ClRe99}. It is most interesting 
to compare the predicted value $(s/\rho_{\rm B})^{\rm RHIC} = 
150 - 170$ to the upcoming RHIC data. 

The evolution of the pressure with the energy density is depicted in 
figure~\ref{fig1}(c). The pressure drops linearly with
the decreasing energy density for all three energies in question. 
Thus, the equation of state in the $(P,\varepsilon)$-plane has a very 
simple form:
$P = 0.12\, \varepsilon$ at AGS, and $P = 0.15\, \varepsilon$ at SPS
and RHIC, i.e. the ratio $P /\varepsilon$ in the central cell is 
saturated already at SPS energies.

Figure~\ref{fig1}(d) presents the evolution of the EOS in the 
$(T, \mu_B)$-plane. In accordance with general estimates
\cite{ClRe99,BMS98} the baryon chemical potential at RHIC energies 
is small while the temperatures are well above the anticipated 
temperature of the QCD phase transition, $T \approx 150$ MeV.
Note that the evolution of temperature 
as a function of ${\varepsilon / \rho_{\rm B}}$ and $\mu_{\rm B}$
is mainly determined by the change of energy per baryon, 
which drops to $\approx 80-90$\% of its initial value,
but not by changing of the baryon chemical potential.  

The slopes of energy spectra of particles in the cell are steeper 
compared to the predictions of the statistical model \cite{rhic}. 
This means that the apparent temperatures of hadrons, especially 
pions, are lower than the temperature given by the SM. 
Since the energy density $\varepsilon$ is the same in both models,
one might expect that the fractions of mesons and resonances in the 
UrQMD cell are overpopulated. These extra-particles consume 
significant part of the total energy and effectively ``cool" the 
hadron cocktail. 

\begin{figure}[htb]
\vspace{-1.5cm}
\hspace{3.cm}
\ewxy{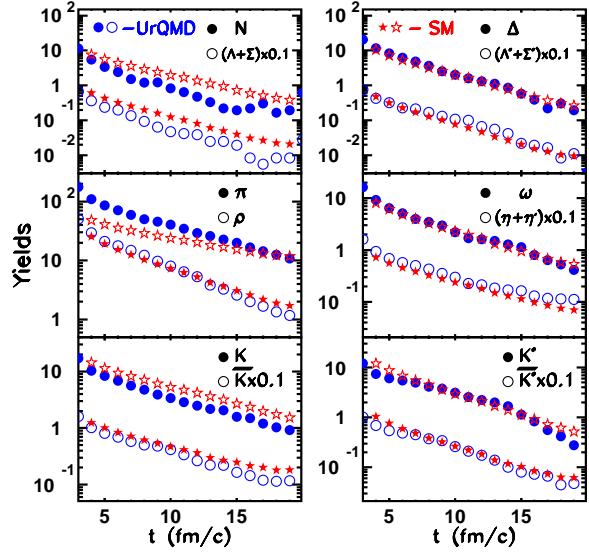}{88mm}  % Yields of hadrons
\vspace{-0.0cm}
\caption{
The yields of main hadron species in the central cell of Au+Au 
collisions at $\sqrt{s} = 200${\it A\/} GeV as a function of time as 
obtained in the model UrQMD (circles) together with the 
predictions of the SM (stars).
}
\vspace{-0.5cm}
\label{fig2}
\end{figure}

Figure \ref{fig2} depicts the yields of the main hadron species 
in the cell within the time interval 5 fm/$c$ $\leq t \leq$ 19
fm/$c$. It is interesting that microscopic spectra 
of pions, which are underestimated by the SM in the central cell
at lower energies \cite{lv99prc}, converge to the SM predictions 
at $t \approx 15$ fm/$c$. Also, the statistical
model overestimates yields of nucleons, lambdas,
and kaons, while the yields of both baryon and meson resonances
are reproduced quite well. Since the baryon number 
and the strangeness are conserved in strong interactions, where is
the rest of the hypercharge, $Y = {\rm B + S}$, in the UrQMD 
calculations in the cell? 
As a matter of fact, the total yields of baryons and antibaryons in
the SM are larger than those of the UrQMD. For instance, at
$t = 10$ fm/$c$ the values of partial baryon and antibaryon density
in the microscopic model are $R_{\rm B}^{\rm mic} = 0.05$ fm$^{-3}$
and $R_{\rm \bar{B}}^{\rm mic} = 0.02$ fm$^{-3}$, respectively, 
while the SM predicts the values $R_{\rm B}^{\rm SM} = 0.08$
fm$^{-3}$ and $R_{\rm \bar{B}}^{\rm SM} = 0.05$ fm$^{-3}$. 
The hadron-resonance-string matter in the cell 
is not in chemical equilibrium; that is why the density of
antibaryons is 2-3 times smaller than the equilibrium values.
Similarly, the SM predicts
significantly larger abundances of both strange baryons and strange
antibaryons at $5 \leq t \leq 11$ fm/$c$ in the cell, while the
densities of strange mesons are nearly the same. 

\begin{figure}[htb]
\vspace{-1.6cm}
\hspace{4.cm}
\ewxy{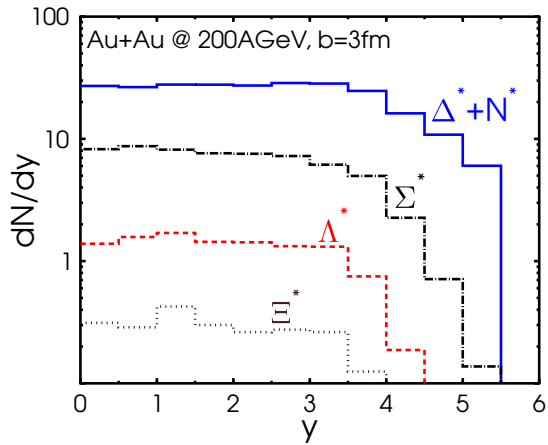}{75mm}  % dN/dy of baryon resonances
\vspace{-0.0cm}
\caption{
The rapidity distributions of baryon resonances in Au+Au
($\sqrt{s}=200${\it A}~GeV) interactions with the impact parameter
$b = 3$ fm.
}
\label{fig3}
\end{figure}

To check the relevance of the central cell results for a 
larger reaction volume at RHIC energies the rapidity distributions 
of baryon resonances are presented in figure~\ref{fig3} (other 
global observables have been studied in \cite{Blei00}). 
Here only those resonances that decay directly into groundstate
hadrons (no final state interactions) are accounted for. These 
distributions are remarkably flat within the 
interval $ |y| \leq 3.5$. More than 80\% of the baryon 
non-strange resonances are $\Delta$'s (1232). The fraction of resonances 
reconstructible from the experimental data is quite large and, 
therefore, the formation 
of resonance-abundant state can be traced experimentally.

\section{Conclusions}
In summary, the following conclusions can be drawn. 
The expansion of matter in central Au+Au collisions at 
$\sqrt{s} = 200${\it A\/} GeV
proceeds with constant entropy per baryon ratio in the central 
cell, $S/A = s/\rho_{\rm B} \cong 150$. Since the $S/A$ ratios for
the central cell in A+A collisions, calculated at AGS and SPS
energies, are very close to the ratios extracted from the analysis
of the experimental data, the expected value of the entropy per
baryon ratio at RHIC lies within the range $150 \leq s/\rho_{\rm B} 
\leq 170$. The microscopic pressure in the cell is close to the 
SM pressure. It shows a linear dependence on the energy density in 
the cell, $P = 0.15\, \varepsilon$.
The obtained result is in accord with the
EOS $P \cong 0.2\, \varepsilon$, derived for an ideal gas of
hadrons and hadron resonances \cite{Shur73}. The temperature 
$T^{\rm SM}$ in the cell at RHIC energies is shown to be nearly
independent on the baryon chemical potential 
$\mu_{\rm B} \approx 50$ MeV.

The comparison of the energy spectra and yields of hadrons with 
those of the SM shows that the times $t \approx 20$\,fm/$c$ may be 
too short to reach full thermal and chemical equilibrium.
Particularly, the deceleration of the relaxation to equilibrium is 
attributed to the creation of the long-lived resonance-abundant 
matter. The yields of resonances are in accord with the SM 
spectra from the very early times $t \approx 5$ fm/$c$,
while the densities of strange and non-strange baryons and their 
antiparticles are lower than the equilibrium values. (However, the 
fitting temperature of the thermal model is higher than the
temperatures of hadron species in the cell.) According to microscopic 
calculations, the resonance-rich matter survives until the thermal 
freeze-out. It remains a challenging task to verify the
formation of long-lived resonance-abundant matter in heavy ion
collisions at $\sqrt{s} = 200${\it A\/} GeV experimentally.

\end{document}